\documentclass[9 pt,journal]{IEEEtran}

\usepackage{booktabs}
\usepackage{tabularx}
\usepackage{array}
\usepackage{graphicx}
\usepackage{hyperref}
\usepackage[cmex10]{amsmath}
\usepackage{amssymb}
\usepackage{url}
\usepackage{titlesec}

\usepackage{amsmath}
\usepackage{subcaption}  

\usepackage{subcaption}  
\usepackage{enumitem}  

\usepackage{array}
\usepackage{adjustbox}

\usepackage{url}
\usepackage[noadjust]{cite}
\usepackage{color}

\usepackage[noadjust]{cite}
\usepackage{color}

\usepackage{algorithm}
\usepackage{algpseudocode}

\usepackage{amssymb}
\usepackage{nomencl}
\makenomenclature

\usepackage{etoolbox}

\titlespacing*{\section}{0pt}{\baselineskip}{0pt}
\titlespacing*{\subsubsection}{0pt}{\baselineskip}{0pt}

\begin{document}

\title{Cyber-Physical Testbed Integrating RTAC with RTDS for Game-Theoretic Topology Control Under Load Altering Attacks}

\author{{Alaa Selim, \emph{Student Member, IEEE}, Junbo Zhao, \emph{Senior Member, IEEE} 

\thanks{
A. Selim and J. Zhao  are with the Department of Electrical and Computer Engineering, University of Connecticut (e-mail: alaa.selim@uconn.edu, junbo@uconn.edu).
}

}}

\markboth{}%
{Shell \MakeLowercase{\textit{et al.}}: Bare Demo of IEEEtran.cls for Journals}

\maketitle


\begin{abstract}
This paper introduces a cyber-physical testbed that integrates the Real-Time Digital Simulator (RTDS) with the Real-Time Automation Controller (RTAC) to enhance cybersecurity in electrical distribution networks. Focused on addressing vulnerabilities to cyber attacks, our testbed employs an advanced control algorithm developed in Python and executed through real-time controllers. Key to our approach is the seamless integration of the host PC machine, RTDS, and RTAC via the Modbus protocol. We present a game theory-based topology control strategy as an effective response to cyber attacks, specifically targeting smart meters. This testbed validates the efficacy of our method in fully eliminating voltage violations due to load altering attacks, showcasing substantial advancements in smart grid cybersecurity via the innovative use of RTDS and RTAC to simulate and counteract complex cyber threats.
\end{abstract}
\vspace{-0.1cm}
\begin{IEEEkeywords}
 Cybersecurity, Cyber-Physical Systems, Game Theory, Cyber Attack Mitigation, Real-Time Digital Simulator (RTDS), Real-Time Automation
Controller (RTAC)

\end{IEEEkeywords}

\vspace{-0.3cm}
\section{Introduction}

\IEEEPARstart{W}{ith} the advent of smart grid technology, the integration of cyber systems with physical grid operations has dramatically transformed the landscape of the power sector. This fusion has ushered in an era of enhanced efficiency and sophisticated management capabilities, as evident in the widespread adoption of advanced metering infrastructure (AMI) and intelligent control systems. As the sector continues to evolve amidst increasing digitalization, the focus shifts towards developing resilient systems to not only withstand cyber threats but also recover swiftly from them. For instance, integrating utility devices like switches and capacitors introduces both complexity and an opportunity for increased grid resilience. By leveraging the synergy between cyber advancements, such as real-time data analytics, and physical grid enhancements, such as adaptive protection schemes, we can innovate and improve grid operations.

\begin{figure}[h]
  \centering
 \includegraphics[width=1.0\columnwidth]{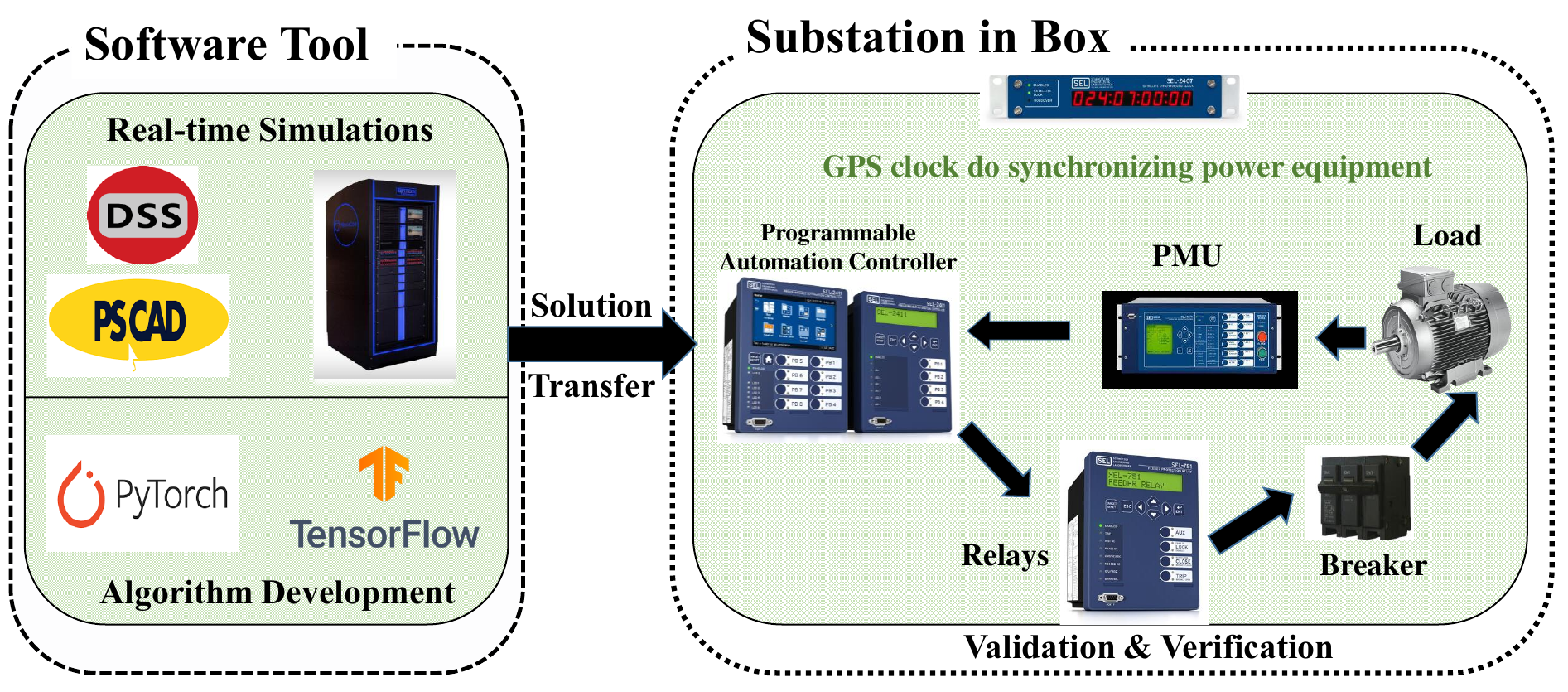}
  \centering
  \caption{Conceptual model}
  \centering
  \label{Fig_coceptual model}
\end{figure}

 The deployment of advanced control algorithms through real-time controllers stands as a cornerstone for validating the effectiveness and feasibility of resiliency solutions. This paper specifically concentrates on harnessing the versatility and power of Python, a high-level programming language known for its efficiency and readability, to address critical power system problems by enabling
the use of advanced machine learning and model-based algorithms \cite{raschka2020machine}. We aim to bridge the gap between theoretical algorithm development and practical application by integrating Python with real-time simulation environments, as shown in Fig. \ref{Fig_coceptual model}.

 Several methodologies are available to achieve seamless integration between Python and real-time controllers, each with its own characteristics and suitability. While API interfaces and socket communication are common methods, they present certain limitations in the context of real-time operations. API interfaces \cite{kornienko2021principles}, for instance, often lack the immediacy required for real-time data exchange, as they typically operate over HTTP protocols not optimized for low latency. Socket communication \cite{xue2009socket}, on the other hand, while potentially faster and more direct, introduces complexities in data packet management and necessitates a more in-depth understanding of network protocols.
In contrast to these methods, Modbus protocol \cite{goldenberg2013accurate} emerges as a more suitable candidate for our integration needs. Despite being slower than socket communication in raw speed, Modbus compensates with its remarkable stability, standardized communication format, and ease of implementation. Its simplicity and robustness, essential for industrial applications, outweigh the speed advantage of socket communication, especially considering the latter's increased complexity and potential for errors in data handling.

\begin{table*}[!ht] 
\centering
\caption{Comparison of Cyber-Physical System Papers in the literature}
\label{table:extended_technical_comparison}
\begin{tabular}{
  >{\centering\arraybackslash}p{3.5cm} 
  | >{\centering\arraybackslash}p{1cm} 
  | >{\centering\arraybackslash}p{1cm} 
  | >{\centering\arraybackslash}p{1cm} 
  | >{\centering\arraybackslash}p{1cm}
  | >{\centering\arraybackslash}p{1cm}
  | >{\centering\arraybackslash}p{1cm}
  | >{\centering\arraybackslash}p{1cm} }
\hline
\textbf{Criteria} & \textbf{\cite{khan2020cyberphysical}} & \textbf{\cite{huang2021real}} & \textbf{\cite{xiao2015development}} & \textbf{\cite{oyewumi2019isaac}} & \textbf{\cite{shariatzadeh2014real}} & \textbf{\cite{biswas2013real}} & \textbf{Our Study} \\  
\hline\hline
Use of RTDS &  & & $\checkmark$ & $\checkmark$ & $\checkmark$ & $\checkmark$  &$\checkmark$  \\
Python Integration with PC &  &  &  &  &  &  &$\checkmark$  \\
Hardware-In-the-Loop & $\checkmark$ & & $\checkmark$ & $\checkmark$ & $\checkmark$ & $\checkmark$  & $\checkmark$ \\
Advanced Algorithms &  &  &$\checkmark$   &  & $\checkmark$ & $\checkmark$  &$\checkmark$  \\
Use of RTAC & $\checkmark$ & $\checkmark$ &  & $\checkmark$ & $\checkmark$ & $\checkmark$ & $\checkmark$  \\
Topology control & $\checkmark$ &  &  &  & $\checkmark$ & $\checkmark$  &$\checkmark$  \\
Use of industrial protocol & $\checkmark$ & $\checkmark$ &  & $\checkmark$ & $\checkmark$ & $\checkmark$  &$\checkmark$  \\
Cyber attack focus & &  &  &  &   &  & $\checkmark$  \\
\hline
\end{tabular}
\label{Table I}
\end{table*}

With the successful integration of Python with real-time controllers such as Real-Time Automation Controller (RTAC) and Real-Time Digital Simulator (RTDS) via Modbus, we are now strategically positioned to tackle pressing challenges in power systems, particularly cyber attacks targeting smart meters. Such attacks pose a significant threat, as they can lead to system instability and violations of operational standards, undermining the reliability and security of the power grid. The sophistication of these cyber attacks requires an equally advanced response – a task well-suited to Python's capabilities. By deploying these Python-developed algorithms on real-time controllers, we can effectively monitor smart meter operations, detect anomalous behaviors indicative of cyber attacks, and initiate immediate countermeasures to maintain system stability and compliance with regulatory standards.

The landscape of smart grid cybersecurity is continually evolving, as highlighted by the comprehensive analysis in \cite{kim2013topology}, which focuses on the complexities of man-in-the-middle attacks altering network topology data. In the context of this expanding field, our research emerges as a crucial next step, concentrating specifically on the cybersecurity of grid meters. Our work diverges from the traditional approach of network data manipulation, instead directing attention toward the vulnerabilities inherent in metering systems and the application of topology control as an innovative mitigation technique. This shift in focus is driven by the need to address the unique challenges posed by meter-centric cyber threats, which have not been extensively explored in existing literature.

In the realm of cyber attack mitigation algorithms, numerous sophisticated methods have been explored, including Deep Reinforcement Learning (DRL) and Model Predictive Control (MPC) as in \cite{selim2023deep}. Recognizing the diverse landscape of these approaches, this paper introduces a novel perspective by implementing a game theory-based approach. Our method focuses on identifying the optimal topology changes to mitigate cyber threats effectively. By analyzing and predicting the outcomes of various scenarios, our algorithm strategically reconfigures grid topology, offering a proactive stance in safeguarding the power system's integrity against cyber vulnerabilities.

In this study, we deploy an advanced cyber-physical testbed to systematically address the complexities of cyber attacks in power systems, leveraging a game theory-based mitigation approach. The efficacy and sophistication of our testbed are critically assessed through a detailed comparative analysis with contemporary cyber-physical testbeds, as delineated in Table \ref{Table I}. Our testbed excels in ease of use and efficient integration of PC-implemented advanced Python algorithms, making it highly effective for cyber attack mitigation in power systems. The main contributions of this paper can be summarized as follows:

\begin{enumerate}
    \item \textbf{Advanced Cyber-Physical Testbed:} The study effectively utilizes a state-of-the-art cyber-physical testbed, showcasing its capabilities in simulating and evaluating game theory-based algorithms against cyber threats.
    \item \textbf{Novel Cyber Attack Scenario:} Introduction of a novel and elusive cyber attack on smart meters, designed to challenge detection methods and cause significant system voltage violations, underscoring the complexity of modern cybersecurity threats.
    \item \textbf{Game Theory-based Topology Control Strategy:} Proposal of a strategic, minimalistic topology control method, leveraging game theory for efficient cyber attack mitigation with reduced switching actions, demonstrating an effective approach to maintaining system stability.
\end{enumerate}

\section{Designing a cyber Attack Vector for Voltage Violations}

\begin{figure}[h]
  \centering
 \includegraphics[width=0.9\columnwidth]{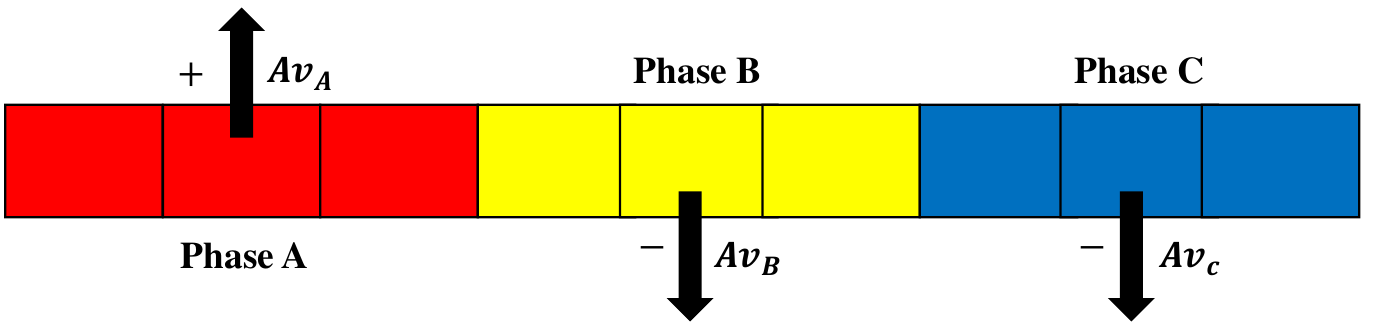}
  \centering
  \caption{Attack vector}
  \centering
  \label{Fig_attack vector}
\end{figure}

Informed by insights from \cite{liu2011false}, we develop the cyber attack model based on a targeted false data injection attack scenario. These attacks are recognized for their stealth, targeting specific state variables to manipulate the power grid's smart meters. The $A\textit{v}$ vector in our model exemplifies this approach by strategically manipulating phase-wise loads to induce maximum voltage violations. This underscores the importance of enhancing security measures in power systems to mitigate such stealthy and targeted attacks. This vector targets the three phases of the system, represented as:
\begin{equation}
    A\textit{v} = \begin{pmatrix} A\textit{v}_A\\ A\textit{v}_B \\ A\textit{v}_C \end{pmatrix}
\end{equation}
where $A\textit{v}_A$, $A\textit{v}_B$, and $A\textit{v}_C$ correspond to phase-wise load alterations. The attack is executed through a controlled manipulation of a selected subset of smart meters, as shown in Fig. \ref{Fig_attack vector}. The methodology involves:
\begin{itemize}
    \item \textbf{Overloading Phase A:} Increasing load in phase A, denoted as $A\textit{v}_A$, to create a high-load scenario.
    \item \textbf{Underloading Phases B and C:} Concurrently reducing loads in phases B ($A\textit{v}_B$) and C ($A\textit{v}_C$) to exacerbate system imbalances and vice versa.
    \item \textbf{Adaptive Monitoring:} Continuous observation of network responses, followed by these adjustments in $A\textit{v}_A$, $A\textit{v}_B$, and $A\textit{v}_C$ for probable cyber attack scenario.
\end{itemize}

The aim is to quickly induce voltage violations by exploiting specific vulnerabilities in the network. The goal of the attack vector is to precipitate the maximum voltage violations with minimal interventions. 
Under constraints, ensure minimal changes in load settings and adherence to network operational standards. This attack model can be represented by an optimization problem as follows:
\begin{align}
    & \max_{\{A\textit{v}_{A,t}, A\textit{v}_{B,t}, A\textit{v}_{C,t}\}_{t=1}^{T}} 
V_{vio,t}  
 \\
    & \text{s.t. } \sum_{t=1}^{T} ( A\textit{v}_{A,t} +  A\textit{v}_{B,t} +  A\textit{v}_{C,t}) \leq  A\textit{v}_{max} \\
    &  A\textit{v}_{A,t} = \begin{cases} 
       A\textit{v}_{A,t-1} \pm \gamma_A, & \text{if } V_{vio,t} < n_{tar} \\
      A\textit{v}_{A,t-1}, & \text{if } V_{vio,t} \geq n_{tar} 
      \end{cases} \\
    &  A\textit{v}_{B,t} = \begin{cases} 
       A\textit{v}_{B,t-1} \pm \gamma_B, & \text{if } V_{vio,t} < n_{tar} \\
      A\textit{v}_{B,t-1}, & \text{if } V_{vio,t} \geq n_{tar} 
      \end{cases} \\
    &  A\textit{v}_{C,t} = \begin{cases} 
       A\textit{v}_{C,t-1} \pm \gamma_C, & \text{if } V_{vio,t} < n_{tar} \\
      A\textit{v}_{C,t-1}, & \text{if } V_{vio,t} \geq n_{tar} 
      \end{cases} \\
    & \gamma(V_{vio}, n_{tar}) = \begin{cases} 
        \alpha, & \text{if } V_{vio} = 0 \\
        \beta(V_{vio}), & \text{if } 0 < V_{vio} < n_{tar}
    \end{cases} \\
    & \beta(V_{vio}) = k - \delta \times V_{vio}, \text{ where } k, \delta > 0 \\
    &  A\textit{v}_{A,t} \in [P_{D_A} \cdot n_{min}, P_{D_A} \cdot n_{max}], \quad \forall t \\
    &  A\textit{v}_{B,t} \in [P_{D_B} \cdot n_{min}, P_{D_B} \cdot n_{max}], \quad \forall t \\
    &  A\textit{v}_{C,t} \in [P_{D_C} \cdot n_{min}, P_{D_C} \cdot n_{max}], \quad \forall t \\
    & n_{min} > 0 \\
    & n_{max} = f(V_{vio} - V_{vio_{init}}) 
\end{align}
where:
\begin{itemize}
    \item \(A\textit{v}_{A,t}, A\textit{v}_{B,t}, A\textit{v}_{C,t}\): Attack vectors on phases A, B, and C at time \(t\).
    \item \(V_{vio,t}\): Number of voltage violations defined per \cite{kusko2007power} standard at time \(t\).
    \item \(A\textit{v}_{max}\): Maximum allowed sum of attack vectors across phases.
    \item \(\gamma_A, \gamma_B, \gamma_C\): Incremental change values for phases A, B, and C.
    \item \(n_{tar}\): Target number of voltage violations.
    \item \(\gamma(V_{vio}, n_{tar})\): Function defining change parameter based on current violations and target.
    \item \(\alpha, \beta(V_{vio})\): Parameters in the \(\gamma\) function, where \(\beta\) is a function of current violations.
    \item \(k, \delta\): Constants in the \(\beta\) function.
    \item \(P_{D_A}, P_{D_B}, P_{D_C}\): Normal load settings for phases A, B, and C.
    \item \(n_{min}, n_{max}\): Minimum and maximum multipliers for load settings.
    \item \(V_{vio_{init}}\): Initial number of voltage violations.
    \item \(f(V_{vio} - V_{vio_{init}})\): Function relating \(n_{max}\) to the change in voltage violations.
\end{itemize}

The cyber attack in our study mimics a voltage unbalance fault, subtly designed to evade detection by standard relays. By causing a slight voltage unbalance, it avoids triggering typical monitoring systems, highlighting the need for more sensitive detection methods to identify such stealthy threats. For this reason, it is essential to calculate the voltage unbalance ($U_t$) for each attack to ascertain that it remains undetectable, thereby ensuring the attack's stealthiness.

\begin{equation}
U_t = \frac{\max\left(\begin{bmatrix} |V_a - V_{avg}| \\ |V_b - V_{avg}| \\ |V_c - V_{avg}| \end{bmatrix}\right)}{V_{avg}} \times 100
\end{equation}

\noindent where $V_a, V_b, \text{ and } V_c$ are the voltages of phases A, B, and C, respectively, and $V_{avg} = \frac{V_a + V_b + V_c}{3}$. As per ANSI C84.1 \cite{kusko2007power}, voltage unbalance exceeding 3\% in distribution networks can lead to inefficient operation and potential damage to electrical equipment. The attack vector tends to stay within allowable voltage imbalance limits, making it less easily detectable, as shown in Algorithm \ref{Algorithm 1}.

\begin{algorithm}
\caption{Cyber Attack Vector for load nodes }
\begin{algorithmic}[1]
    \State Initialize network parameters and set initial loads $A\textit{v}_{A,0}, A\textit{v}_{B,0}, A\textit{v}_{C,0}$
    \For{each time step $t = 1$ to $T$}
        \State Adjust load on Phase A: $A\textit{v}_{A,t} = A\textit{v}_{A,t-1} \pm \gamma_A$
        \State Adjust loads on Phase B: $A\textit{v}_{B,t} = A\textit{v}_{B,t-1} \mp \gamma_B$
        \State Adjust loads on Phase C: $A\textit{v}_{C,t} = A\textit{v}_{C,t-1} \mp \gamma_C$
        \State Compute node voltages: $V_{nodes,t}$
        \State Check voltage unbalance factor $U_{t}$ at time $t$
        \State Count voltage violations at time $t$
    \EndFor
    \State Analyze overall attack impact based on voltage violations and unbalance factors

\end{algorithmic}
\label{Algorithm 1}
\end{algorithm}

\section{Game Theory for Attack Mitigation}

In response to the cyber attacks initiated in section II, particularly those targeting network stability and voltage profiles, we propose a game-theoretic approach shown in Algorithm \ref{Algorithm 2} to control network topology. This method involves adjusting switch states within the network to maintain stability, ensure service continuity, and minimize switching actions for practical real-time applications. 

The network consists of a set of switches $S = \{s_1, s_2, \ldots, s_n\}$. Each switch $s_i$ can be in one of two states, represented by the strategy set $S_i = \{\text{open}, \text{closed}\}$. The current state of the network is denoted by $V$, reflecting the voltage profile post-attack as determined by Algorithm \ref{Algorithm 1}.

The key objective is to modify the switch states in a way that ensures the radiality of the network, denoted by $R(S)$. Radiality is a constraint that preserves the tree structure of the network, which is essential for minimizing disruptions and maintaining operational integrity. Additionally, we aim to minimize the number of switching actions, which is crucial for the feasibility of real-time operations. To achieve this, we introduce a cost function \( C(S) \) associated with switch operations.

The payoff function $U(S_i)$ is defined as $f(R(S), V(S), C(S))$, which evaluates the effectiveness of each strategy in terms of maintaining radiality, reducing voltage violations, and minimizing the cost associated with switch operations. The optimal switch state $s_i^{*}$ maximizes the payoff, given by:
\[ s_i^{*} = \arg\max U(S_i) \]
Subject to the constraints of radiality, voltage stability, and minimal switching cost. Given a set of nodes in an electrical network, let \( V_i \) be the voltage at node \( i \), and \( S_{ij} \) represent the state of the switch between nodes \( i \) and \( j \) (1 if closed, 0 if open). The radiality of the network can be mathematically expressed as:

\[
R = \prod_{i,j \in \text{Nodes}} \left( 1 - \delta(V_i, 0) \delta(V_j, 0) (1 - S_{ij}) \right)
\]

\noindent where \( \delta(x,y) \) is the Kronecker delta function. This function equals 1 if \( x = y \) and 0 otherwise. The radiality metric \( R \) will be 1 if the network maintains radiality, and 0 if it doesn't. The switching cost function \( C(S) \) is designed to penalize frequent switching, thereby encouraging solutions that maintain stability with minimal switch operations.

\begin{algorithm}
\caption{Game-Theoretic Topology Control for Attack Mitigation with Minimized Switching}
\begin{algorithmic}[1]
\State Start: Voltage profile from Algorithm 1 after attack
\State Initialize: Switch states $S = \{s_1, s_2, \ldots, s_n\}$
\While{true}
    \State Observe current $V$ and network topology
    \State Ensure radiality $R(S)$ 
    \For{each switch $s_i$ in $S$}
        \State Define strategies $S_i = \{\text{open}, \text{closed}\}$
        \State Calculate payoff $U(S_i) = f(R(S), V(S), C(S))$
        \State $s_i^{*} \leftarrow \arg\max U(S_i)$
    \EndFor
    \State Update switch states $S = \{s_1^{*}, s_2^{*}, \ldots, s_n^{*}\}$ with minimal changes
    \State Apply $S$ to the network
    \State Sleep for control interval
\EndWhile
\end{algorithmic}
\label{Algorithm 2}
\end{algorithm}

\section{Cyber-Physical Testbed Setup}
 We have developed a comprehensive cyber-physical testbed designed to implement our game-theoretic algorithm for mitigating cyber attacks. This setup involves executing Python code on a PC machine, with subsequent communication to RTDS through a GTNETx2 card. The integration of the PC machine with RTDS is achieved via two distinct approaches.
\subsubsection{Direct Python-to-RTDS Integration}
\begin{figure}[h]
  \centering
 \includegraphics[width=1.0\columnwidth]{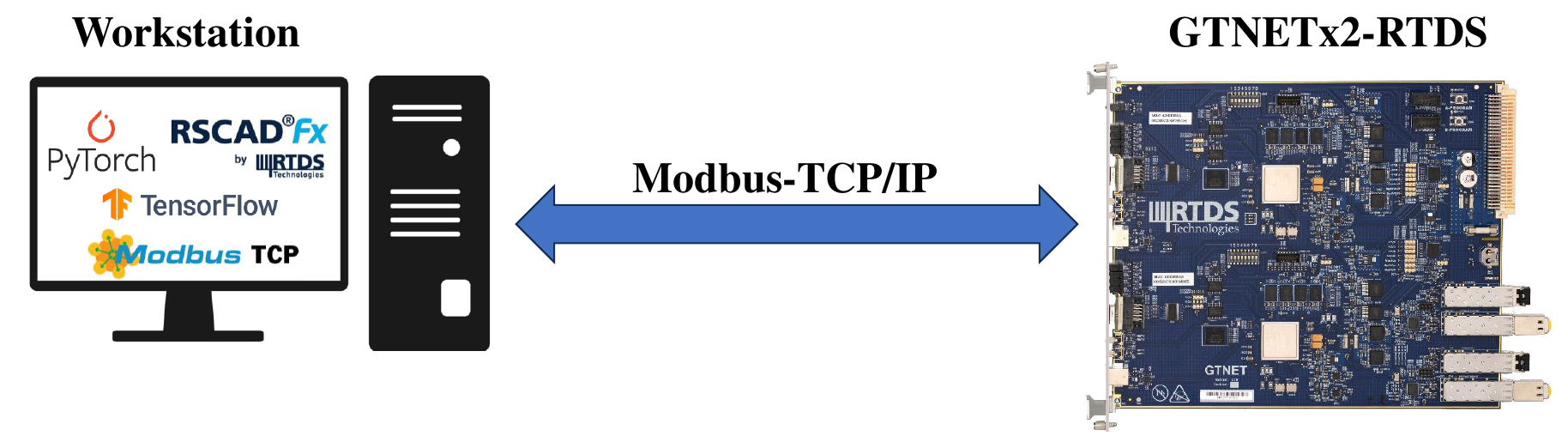}
  \centering
  \caption{Network diagram for research purpose}
  \centering
  \label{Fig_network research}
\end{figure}

The first approach involves directly running Python scripts on the RTDS from the host machine, as shown in Fig. \ref{Fig_network research}. This method is particularly suitable for research purposes due to its straightforward integration. Utilizing the Pymodbus \cite{pymodbus} library, the host machine functions as a client, sending control commands to the RTDS. In this setup, the GTNETx2 card acts as a server, receiving control actions to operate network switches and monitor voltage levels within the network. Data on these voltage levels is then relayed back to the host machine.
\subsubsection{RTAC-Based Integration for Industrial Application}
\begin{figure}[h]
  \centering
 \includegraphics[width=1.0\columnwidth]{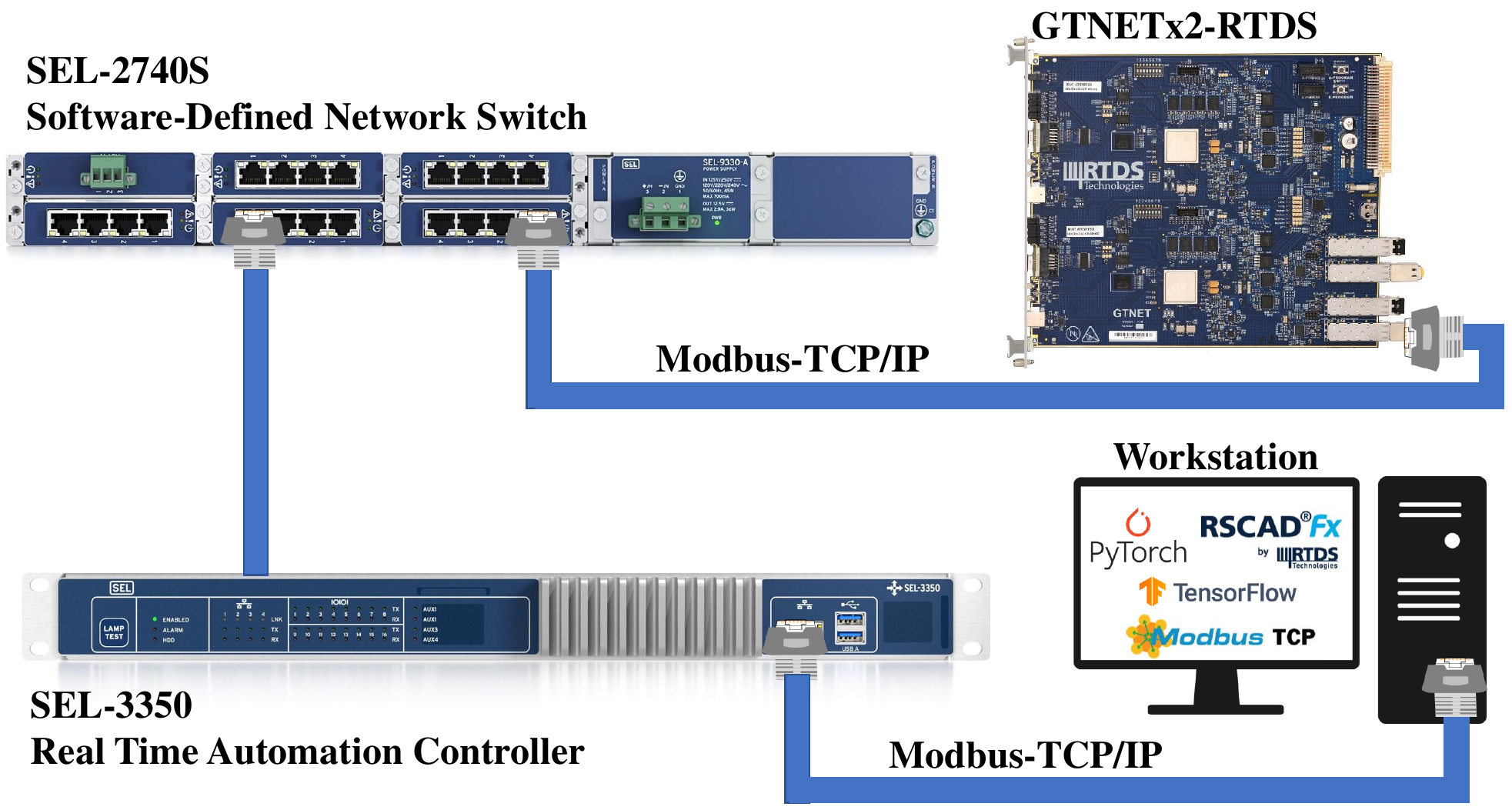}
  \centering
  \caption{Network diagram for deployment in field}
  \centering
  \label{Fig_network field}
\end{figure}

The second approach shown in Fig. \ref{Fig_network field} represents a more industrial application. In this method, RTAC serves as a server to the host machine. The host machine communicates control actions to the RTAC, which then relays this data to the RTDS, acting as a client. This communication is facilitated through a Software-Defined Networking (SDN) infrastructure and network switches. This approach demonstrates the practical application of RTAC in the field, enabling the execution of advanced control actions in real-world scenarios.
Both methods provide valuable insights into the practical implementation of our algorithm. While the first approach offers ease of use and quick integration for research and development, the second approach showcases the potential for deploying such advanced control strategies in an industrial setting, leveraging RTAC's capabilities.

SDN in \cite{brugman2019cloud} significantly enhances network scalability, centralizes control, improves resilience, and increases visibility, making it ideal for managing extensive control actions. The scalability of SDN allows efficient traffic management and inspection, which is crucial for large-scale networks. Its centralized control architecture enables dynamic and efficient packet routing, which is essential for handling numerous control actions. These features collectively make SDN an excellent choice for complex network management demands. 
While integrating Modbus with the RTDS provides numerous advantages, it also presents limitations that must be addressed for effective implementation. Firstly, Modbus supports a maximum of 125 holding registers for reading data. This restriction can be particularly challenging in large-scale applications, as it limits the amount of data retrieved in a single query, potentially leading to incomplete data acquisition. To circumvent this limitation, we have implemented a systematic approach to selectively retrieve the most crucial data, ensuring that critical information is noticed despite the limited number of registers.
Another limitation is that Modbus in RTDS natively supports only 16-bit unsigned integers. This constraint becomes apparent when reading voltage levels, typically represented as floating-point numbers. We employed a technique based on the IEEE 754 Standard for Floating-Point Arithmetic to resolve this. According to this standard, a 32-bit single-precision floating-point number (FLOAT32) can be effectively represented using two consecutive 16-bit registers in the Modbus data model. This mapping allows for the accurate reading of voltage levels as exact float values. Notably, this approach is also recommended by the RTDS support team, aligning with industry best practices for data accuracy and representation in digital systems. This testbed configuration allows for accurate sensor measurements in large-scale applications, ensuring reliable data analysis and control strategies.

\section{Numerical Testing Analysis}
\begin{figure}[h]
  \centering
 \includegraphics[width=1.0\columnwidth]{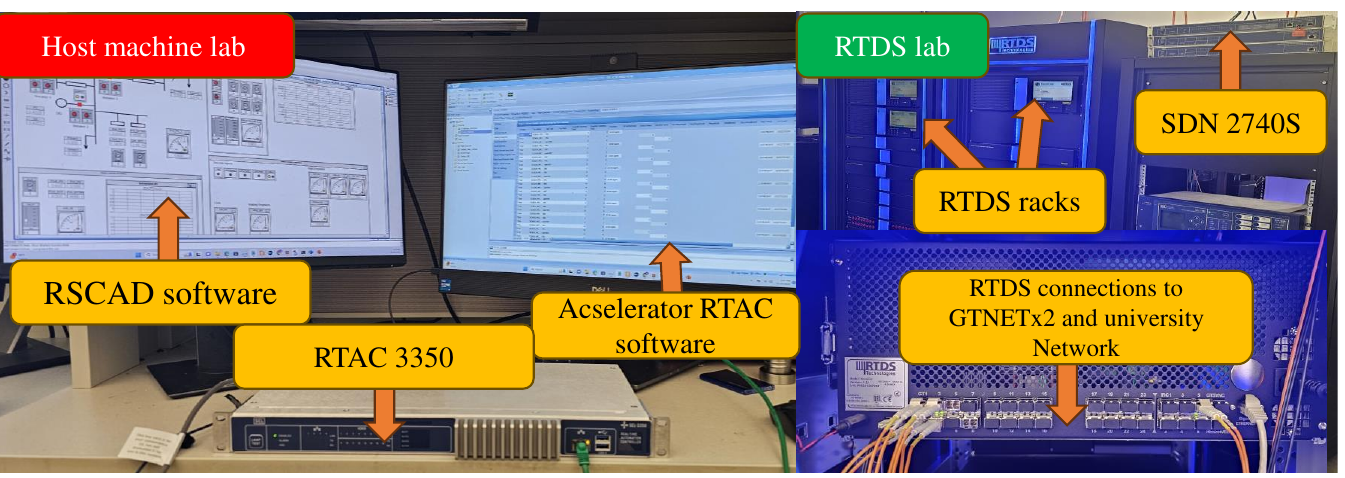}
  \centering
  \caption{Cyber-Physical testbed setup}
  \centering
  \label{Fig_setup}
\end{figure}

\begin{figure}[h]
  \centering
 \includegraphics[width=1.0\columnwidth]{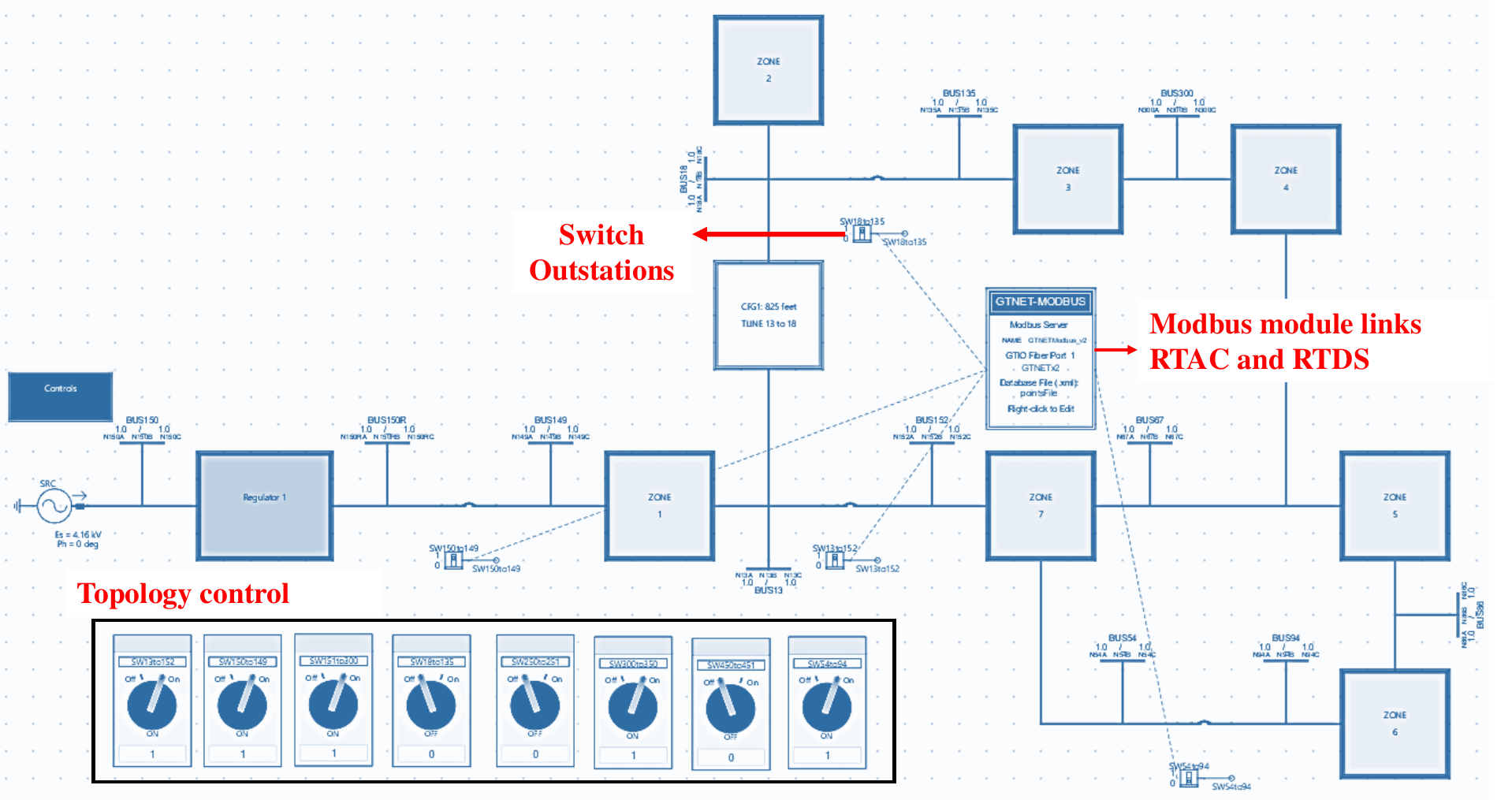}
  \centering
  \caption{IEEE-123 draft in RSCAD}
  \centering
  \label{Fig_draft}
\end{figure}

\begin{figure}[h]
  \centering
 \includegraphics[width=1.0\columnwidth]{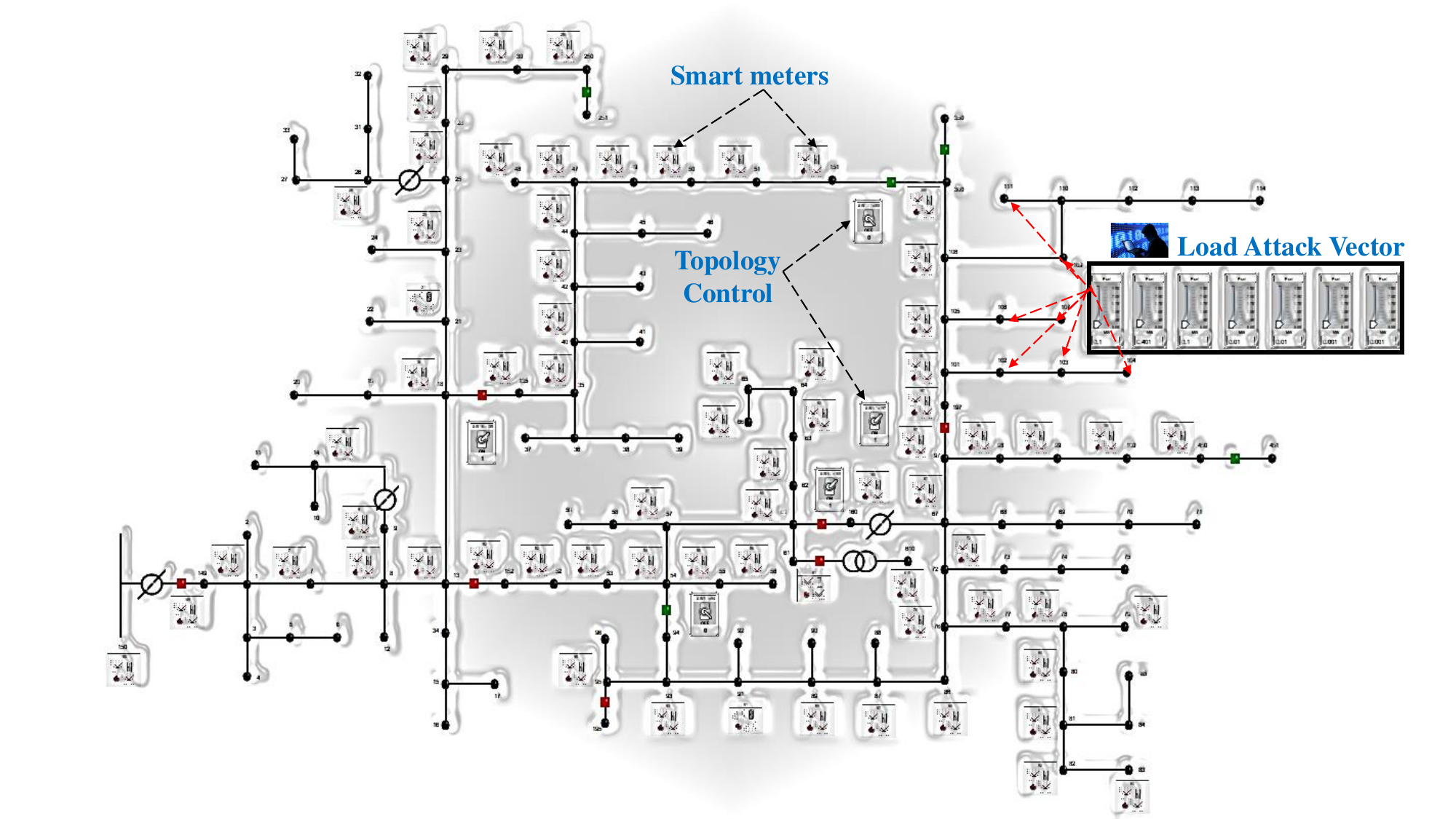}
  \centering
  \caption{IEEE-123 Runtime in RSCAD}
  \centering
  \label{Fig_runtime}
\end{figure}

In our experiment shown in Fig. \ref{Fig_setup}, we conduct tests on the IEEE-123 bus system, implemented using RSCAD version FX2.1 (see Fig. \ref{Fig_draft}), where the system starts typically with zero voltage violations. We employ the Pymodbus library for communication between the host machine and the RTDS and RTAC through Modbus/TCP. The ACSELERATOR RTAC software facilitates the configuration of the RTAC, including mapping of data points and verifying connectivity. Our setup begins with sending control actions to the switches, utilizing the two approaches discussed in the previous section. For data collection, we assume visibility over 206 smart meters in our test setup, as shown in Fig. \ref{Fig_runtime}. However, due to Modbus's limitation of 125 tags in RTDS, we develop a script in RSCAD that prints all voltage readings as soon as the host machine triggers any control action. These voltage data are then saved on the same host machine. This setup allows for the seamless integration of real-time voltage data into our Python-based mitigation algorithm, enabling it to respond dynamically to changes in the network’s state and effectively mitigate any identified voltage violations. Table \ref{tab:modbus_data_exchange} succinctly categorizes the Modbus communication within a cyber-physical system. It details three key components: 'Voltage Data' with Holding Registers 1-206 for individual node voltages, 'Topology Control' using Coils 1-6 for switch operations (S1-S6), and 'Load Control' through Holding Registers 207-215 for various setpoint adjustments. This structure demonstrates the protocol's integral role in managing essential aspects of voltage monitoring, network topology, and load management.

\begin{table}[ht]
\centering
\caption{Modbus Data Exchange with RTAC}
\begin{tabular}{|c|c|c|}
\hline
\textbf{Modbus Entity} & \textbf{Type} & \textbf{Description} \\
\hline
Holding Register 1 & Voltage Data & Node 1 Voltage \\
Holding Register 2 & Voltage Data & Node 2 Voltage \\
\(\vdots\) & \(\vdots\) & \(\vdots\) \\
Holding Register 206 & Voltage Data & Node 206 Voltage \\
\hline
Coil 1 & Topology Control & Switch Control S1 \\
Coil 2 & Topology Control & Switch Control S2 \\
\(\vdots\) & \(\vdots\) & \(\vdots\) \\
Coil 6 & Topology Control & Switch Control S6 \\
\hline
Holding Register 207 & Load Control & Setpoint 1 \\
\(\vdots\) & \(\vdots\) & \(\vdots\) \\
Holding Register 215 & Load Control & Setpoint 9 \\
\hline
\end{tabular}
\label{tab:modbus_data_exchange}
\end{table}

We initiate the cyber attack as outlined in Algorithm \ref{Algorithm 1}, strategically implementing the attack vectors to analyze their impact on the IEEE-123 bus system, starting from zero voltage violations level. The results of our cyber attack simulation, as detailed in Table \ref{tab:attack_vector_impact}, exhibit the nuanced effects of different attack vectors on the power network's voltage profile, highlighting the stealthiness of the attack strategy. In Case 1, the attack vector \( {Av}_C \) increases to 0.08 at nodes N102, N103, and N104, doubling the normal load, while other phases reduce to a mere 0.001. This results in 26 voltage violations with an unbalance of 1.704\%, indicating a significant yet covert disruption, as the unbalance stays below typical detection thresholds. Case 2 mirrors this strategy with attack vector \( {Av}_B \), again leading to 26 violations but with a slightly higher unbalance of 1.888\%. This subtle increase in unbalance, while still remaining undetectable by standard systems, hints at the attack's evolving sophistication. Case 3 focuses on Phase C with similar intensity, yet it results in only three violations, albeit with a higher voltage unbalance of 2.000\%. The reduced number of violations contrasts with the increased voltage unbalance percentage, showcasing the attack's ability to impact network stability without triggering immediate alarms.

\begin{figure*}[ht]
  \centering
  \begin{subfigure}{.32\textwidth}
    \centering
    \includegraphics[width=\linewidth]{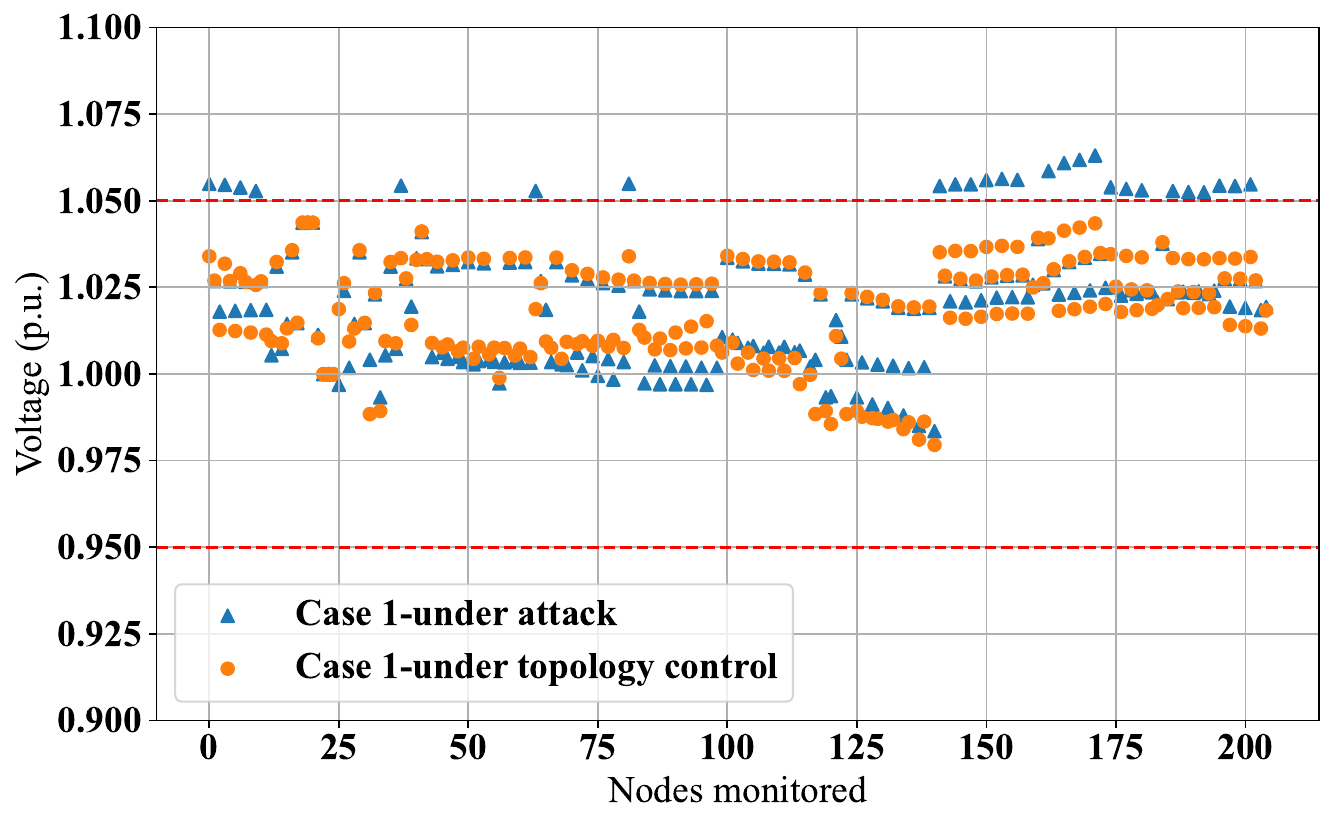}
    \caption{Case 1}
    \label{fig:case1}
  \end{subfigure}%
  \begin{subfigure}{.32\textwidth}
    \centering
    \includegraphics[width=\linewidth]{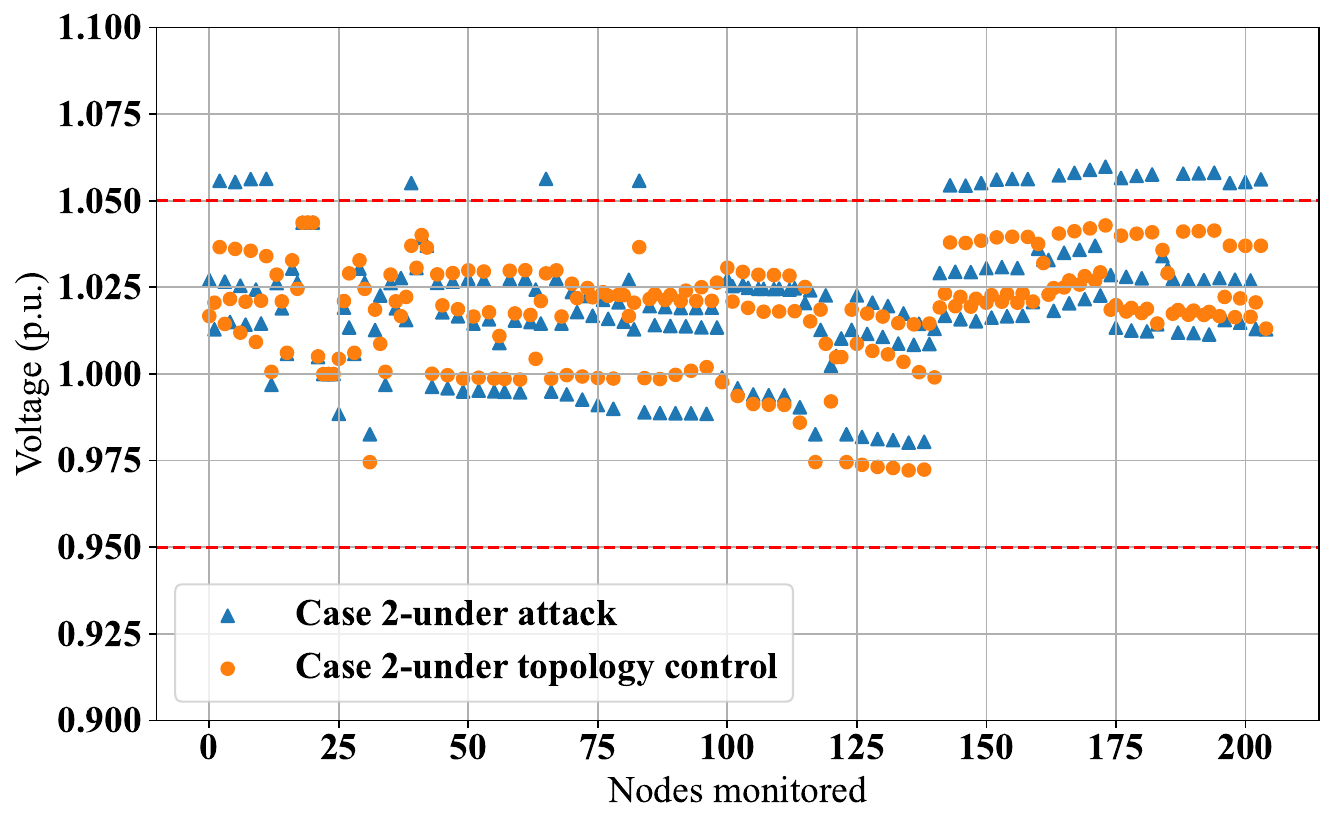}
    \caption{Case 2}
    \label{fig:case2}
  \end{subfigure}%
  \begin{subfigure}{.32\textwidth}
    \centering
    \includegraphics[width=\linewidth]{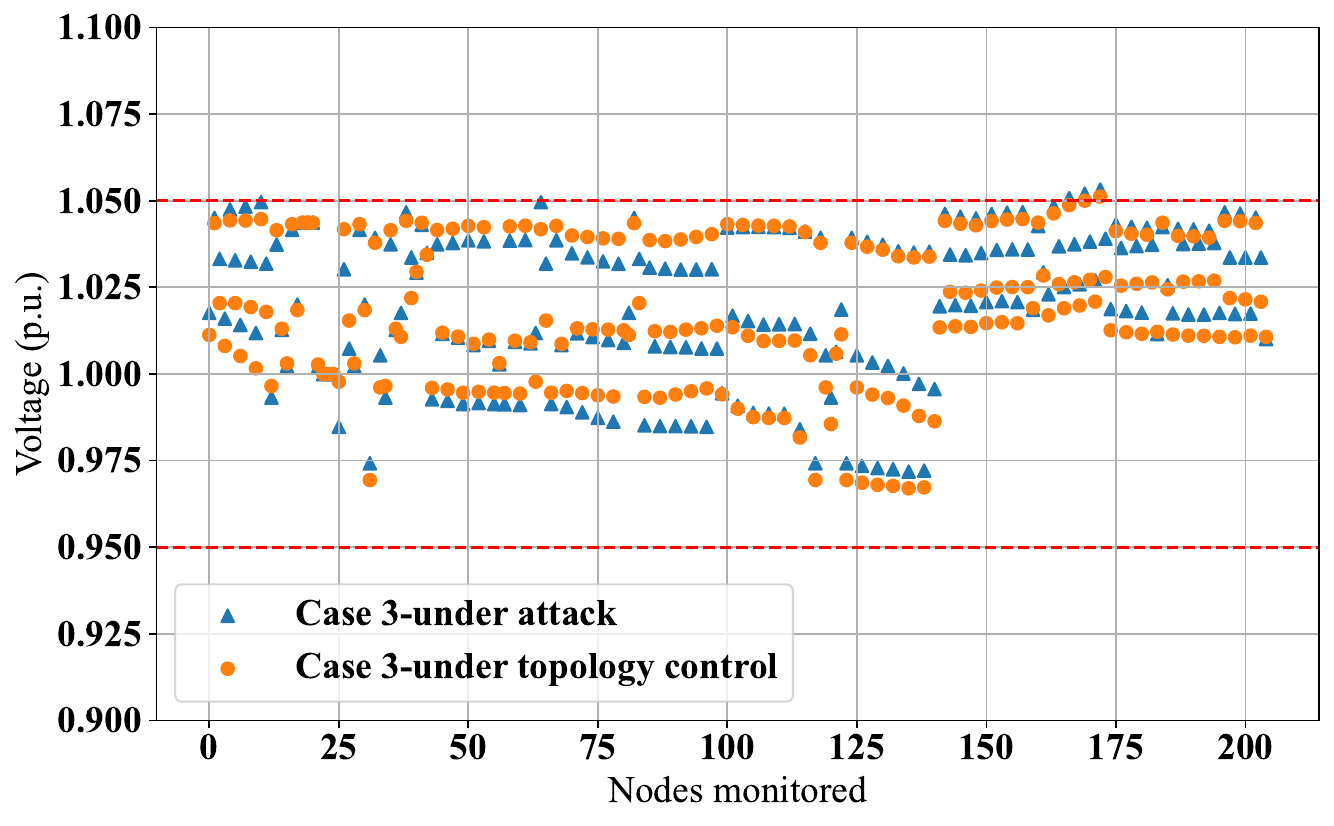}
    \caption{Case 3}
    \label{fig:case3}
  \end{subfigure}

  \begin{subfigure}{.32\textwidth}
    \centering
    \includegraphics[width=\linewidth]{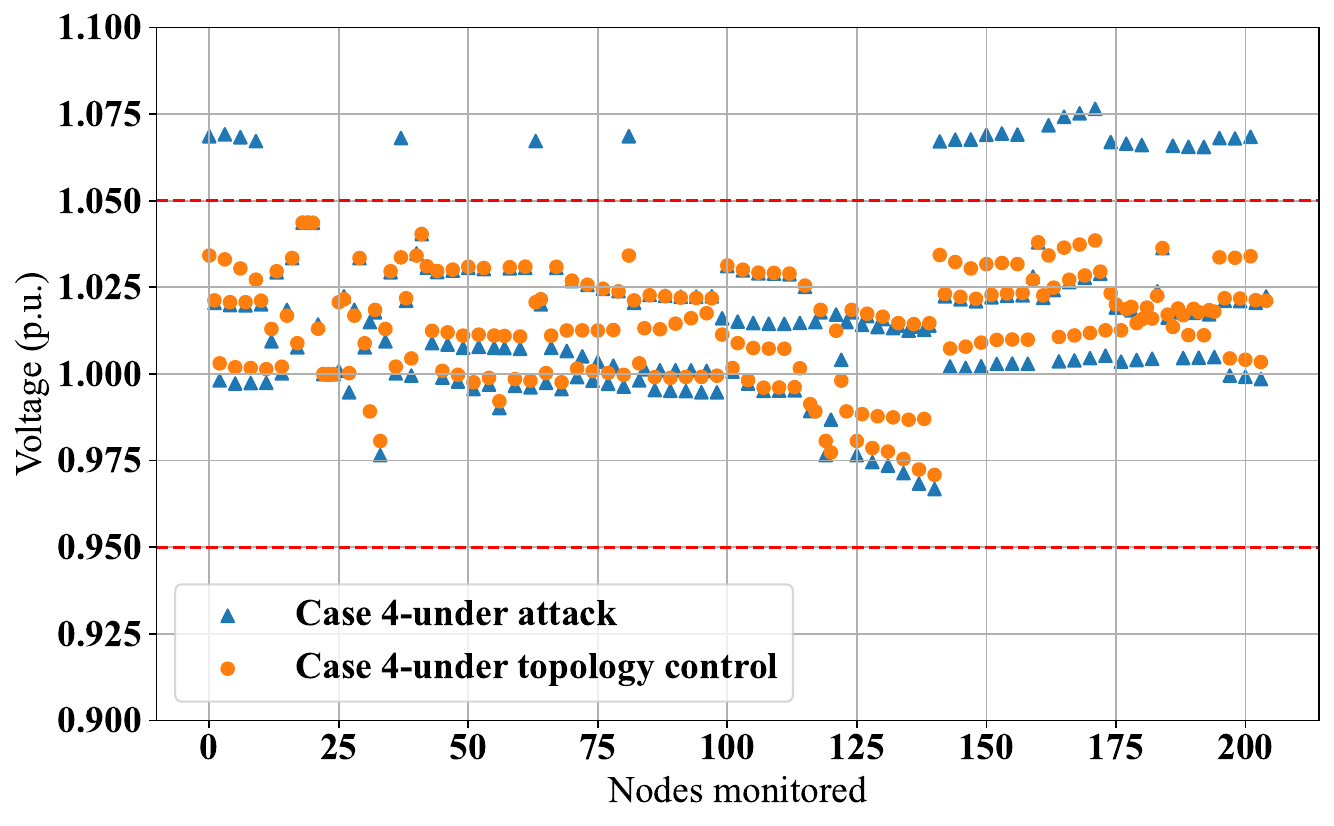}
    \caption{Case 4}
    \label{fig:case4}
  \end{subfigure}%
  \begin{subfigure}{.32\textwidth}
    \centering
    \includegraphics[width=\linewidth]{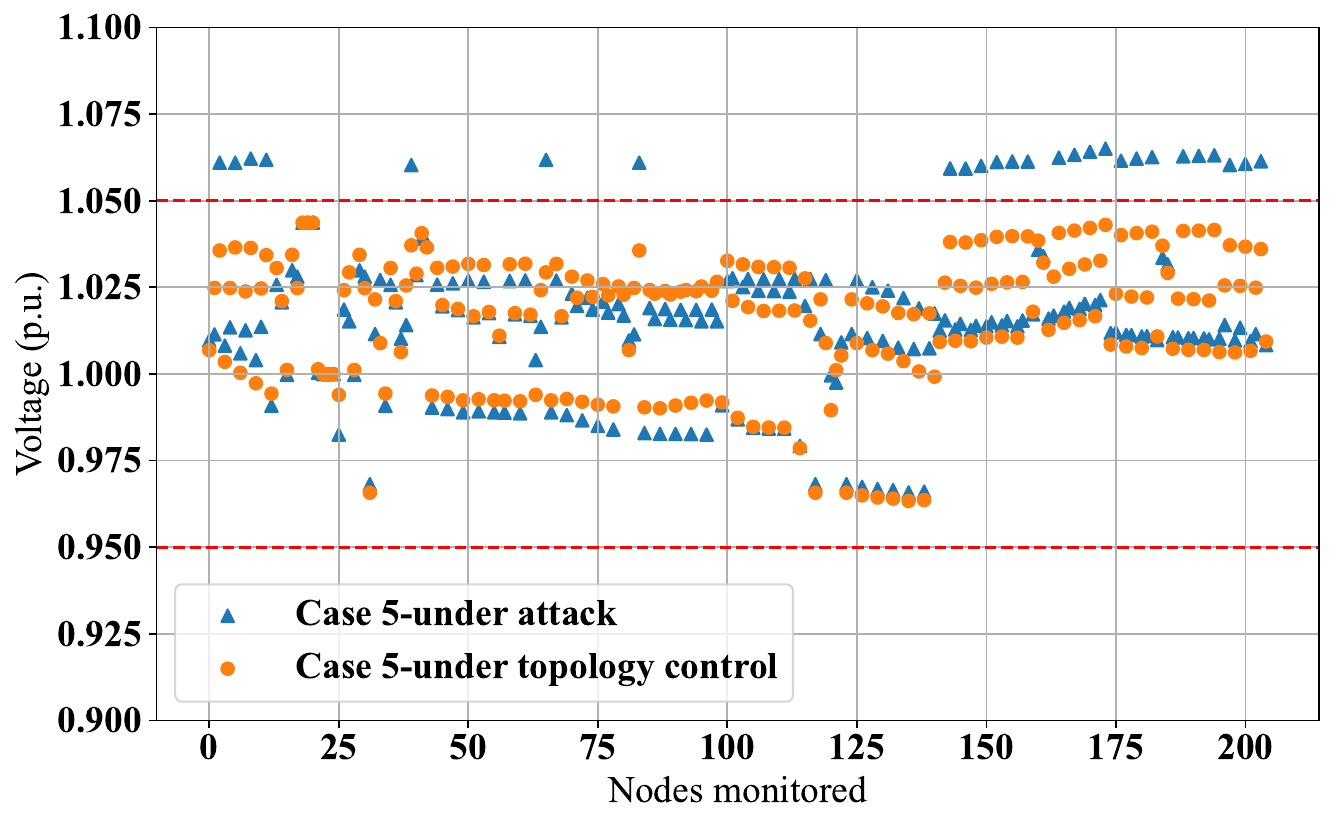}
    \caption{Case 5}
    \label{fig:case5}
  \end{subfigure}%
  \begin{subfigure}{.32\textwidth}
    \centering
    \includegraphics[width=\linewidth]{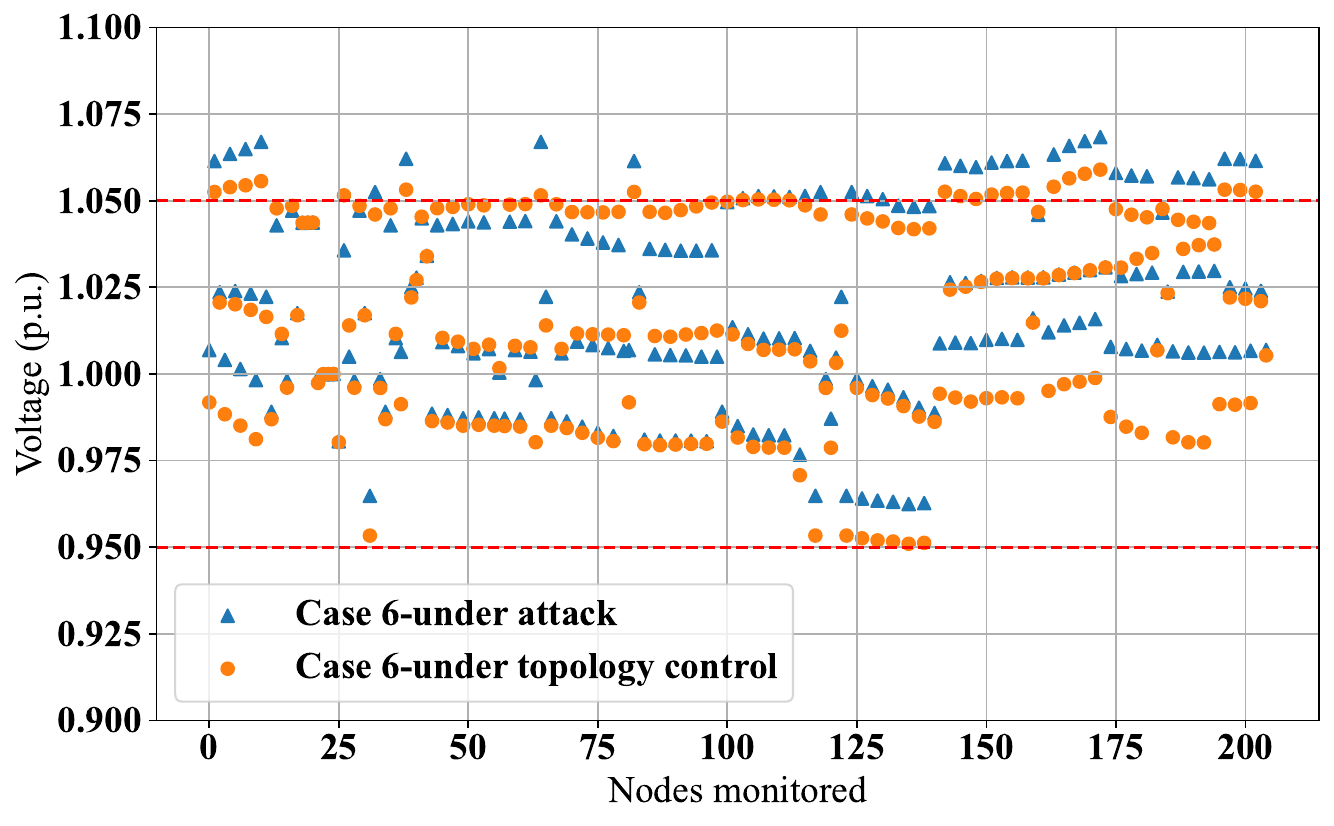}
    \caption{Case 6}
    \label{fig:case6}
  \end{subfigure}
  \caption{Mitigation of voltage violations via topology control in response to load altering attacks }
  \label{fig:voltage_comparisons}
\end{figure*}

In Cases 4, 5, and 6, the attack intensity escalates further to 0.16, significantly overloading one phase while reducing others to near switch-off levels. These cases demonstrate the attack's increasing severity: Case 4 results in 26 violations with a 2.516\% unbalance, Case 5 has the same number of violations, but a slightly higher unbalance of 2.574\%, and Case 6 marks the peak of the attack's impact with 36 violations and a 2.999\% unbalance. These latter cases show a clear departure from the stealthiness observed in earlier scenarios, as the magnitude of the voltage violations and unbalance soar beyond standard limits, posing a severe and overt threat to the network's stability. Fig. \ref{fig:voltage_comparisons} vividly illustrates the magnitude of voltage violations resulting from the attack vector.


In the IEEE-123 bus system, switches S1 to S6 are normally closed (on), ensuring a continuous power flow. Conversely, switches S7 and S8 are typically open (off), allowing for flexible network management and control within the system. The efficacy of post-attack topology control measures, implemented through a game-theoretic approach as detailed in Table \ref{tab:topology_changes}, is clearly demonstrated in their ability to effectively mitigate voltage violations. In Cases 1 through 5, the strategic operation of switches, specifically S7 and S8, successfully eliminates all voltage violations, highlighting the approach's precision in maintaining network stability. Notably, the game-theoretic method excels in optimizing switch states with minimal changes - a crucial factor for real-time operation feasibility. In most cases, this results in maintaining S8 in the 'Off' state while selectively operating S7, thus minimizing the switching actions and preserving the radiality of the system.

Case 6, however, presents a more complex scenario; despite applying similar switch adjustments, with both S7 and S8 'On', the network experiences 25 instead of 36 violations. In addition, the game theory approach explores various switch combinations for this scenario, but it ultimately determines that the current configuration is the most optimal for voltage regulation. This particular outcome underscores the challenge posed by varying attack intensities and the necessity for a more nuanced control strategy. The game-theoretic approach's adaptability and dynamic response are essential in these scenarios, allowing for real-time adjustments to the network's topology. This adaptability not only ensures the mitigation of voltage violations but also maintains the essential radial structure of the power system. Furthermore, Fig. \ref{fig:voltage_comparisons} shows a significant reduction in voltage violations, with the impacted nodes now comfortably within voltage standard limits.

In our study, adjusting switches S7 and S8 effectively mitigates voltage violations from cyber attacks on smart meters. While this highlights their importance in network stability, it is important to recognize that this strategy's effectiveness may vary in different scenarios.

\begin{table}[ht]
\centering
\caption{Impact of Topology Changes on Voltage Violations}
\label{tab:topology_changes}
\resizebox{\columnwidth}{!}{%
\begin{tabular}{|c|c|c|c|c|c|c|c|c|c|}
\hline
\textbf{Case} & \textbf{S1} & \textbf{S2} & \textbf{S3} & \textbf{S4} & \textbf{S5} & \textbf{S6} & \textbf{S7} & \textbf{S8} & \textbf{Violations Count} \\ \hline
1 & On &  On & On &  On & On &  On & \textbf{On} & Off & 0 \\ \hline
2 & On &  On & On &  On & On &  On & \textbf{On} & Off & 0 \\ \hline
3 & On &  On & On &  On & On &  On & \textbf{On} & Off & 0 \\ \hline
4 & On &  On & On &  On & On &  On & \textbf{On} & \textbf{On} & 0\\ \hline
5 & On &  On & On &  On & On &  On & \textbf{On} & Off & 0 \\ \hline
6 & On &  On & On &  On & On &  On & \textbf{On} & \textbf{On} & 25 \\ \hline
\end{tabular}
}
\end{table}

\begin{table*}[htb]
\centering
\caption{Impact of Attack Vectors on Voltage Violations and Unbalance}
\begin{tabularx}{\textwidth}{@{}X *{10}{>{\centering\arraybackslash}X} c@{}}
\toprule
\textbf{Case} & \multicolumn{3}{c}{\textbf{Attack Vector \( {Av}_C \)}} & \multicolumn{3}{c}{\textbf{Attack Vector \( {Av}_B \)}} & \multicolumn{3}{c}{\textbf{Attack Vector \( {Av}_A \)}} & \textbf{Violations} & \textbf{Voltage Unbalance (\%)} \\
\cmidrule(r){2-4} \cmidrule(lr){5-7} \cmidrule(l){8-10}
 & N102 & N103 & N104 & N106 & N107 & N99 & N109 & N111 & N114 & & \\
\midrule
1 & 0.08 & 0.08 & 0.08 & 0.001 & 0.001 & 0.001 & 0.001 & 0.001 & 0.001 & 26 & 1.704 \\
2 & 0.001 & 0.001 & 0.001 & 0.08 & 0.08 & 0.08 & 0.001 & 0.001 & 0.001 & 26 & 1.888 \\
3 & 0.001 & 0.001 & 0.001 & 0.001 & 0.001 & 0.001 & 0.08 & 0.08 & 0.08 & 3 & 2.000\\
4 & 0.16 & 0.16 & 0.16 & 0.001 & 0.001 & 0.001 & 0.001 & 0.001 & 0.001 & 26 & 2.516 \\
5 & 0.001 & 0.001 & 0.001 & 0.16 & 0.16 & 0.16 & 0.001 & 0.001 & 0.001 & 26 & 2.574 \\
6 & 0.001 & 0.001 & 0.001 & 0.001 & 0.001 & 0.001 & 0.16 & 0.16 & 0.16 & 36 & 2.999 \\
\bottomrule
\end{tabularx}
\label{tab:attack_vector_impact}
\end{table*}

\section{Conclusion and Future Work}

This study successfully demonstrated the integration of Python with RTDS and RTAC, creating a robust cyber-physical testbed for real-time power system simulations. We designed and executed a stealthy cyber attack on smart meters, emphasizing the importance of advanced detection and mitigation strategies in modern power systems. Our game-theoretic approach for topology control proved effective in maintaining network stability and minimizing voltage violations, even under challenging attack scenarios.

 Looking ahead, our future work will concentrate on expanding the scope and capabilities of our testbed. This includes enhancing data handling to manage more extensive networks and addressing the limitations of Modbus for more comprehensive data acquisition by using other industrial protocols like DNP3 or Ethernet/IP. We also aim to explore the integration of distributed energy resources (DERs) control alongside topology control in our cyber attack mitigation strategies. By incorporating machine learning algorithms, we aspire to develop more dynamic and predictive control strategies that adapt proactively to the evolving landscape of cyber threats.

\section{Acknowledgment}
The authors are thankful to Schweitzer Engineering Lab (SEL)
and RTDS for their support in successfully implementing this project.

\bibliographystyle{ieeetr}
\bibliography{Mendeley.bib}
\end{document}